\documentclass[journal, twocolumn, letterpaper, twoside]{IEEEtran}

\usepackage{times}
\usepackage{fancyhdr}
\usepackage{lastpage}
\usepackage{amsmath,amsfonts,amssymb,amsthm,amscd,array,mathrsfs,pifont}
\usepackage{subfigure}
\usepackage{color}
\usepackage{cite}
\usepackage{psfrag}
\usepackage{graphicx}
\usepackage{units}
\usepackage{boxedminipage}

\begin{document}
\pagestyle{fancy}
\bstctlcite{IEEEtranBST:BSTcontrol}

\rfoot{\footnotesize DRAFT}
\lfoot{\footnotesize April 7, 2010}  
\chead{\scriptsize Accepted for IEEE Transactions on Circuits and Systems II Express Briefs, May 2010. This draft will not be updated any more. Please refer to
     IEEE Xplore for the final version.
     ${}^{*)}$ The final publication will appear with the
     modified title
    \emph{``A Scalable VLSI Architecture for Soft-Input Soft-Output Single Tree-Search Sphere Decoding''.}
}

\sloppy
\hyphenation{hypo-the-sis ex-trin-sic con-ver-gence}

%
%
%
%

\definecolor{gray}{rgb}{0.3, 0.3, 0.3}
\definecolor{colorMA}{rgb}{0, 0, 1}
\definecolor{colorMC}{rgb}{0, 0.4, 0}

\newcommand{\counterHypothesis}{counter-hypothesis}
\newcommand{\colwise}{column-wise}
\newcommand{\zigzag}{zig-zag}
\newcommand{\apriori}{a priori}
\newcommand{\APriori}{A Priori}
\newcommand{\Apriori}{A priori}
\newcommand{\aprioribased}{a priori-based}
\newcommand{\APrioriBased}{A Priori-Based}
\newcommand{\Aprioribased}{A priori-based}
\newcommand{\hardinput}{\fixmeInvalidTerm}
\newcommand{\HardInput}{\fixmeInvalidTerm}
\newcommand{\softinput}{soft-input}
\newcommand{\SoftInput}{Soft-Input}
\newcommand{\hardoutput}{hard-output}
\newcommand{\softoutput}{soft-output}
\newcommand{\SoftOutput}{Soft-Output}
\newcommand{\softoutputOnly}{soft-output-only}
\newcommand{\SoftOutputOnly}{Soft-Output-Only}
\newcommand{\subtree}{sub-tree}
\newcommand{\STSSD}{STS SD}
\newcommand{\SISOSTSSD}{SISO \STSSD{}}
\newcommand{\mathtext}[1]{\textrm{\normalfont #1}}
\newcommand{\probab}[1]{\mathtext{P}[#1]}
\newcommand{\matrixSym}[1]{\textbf{\normalfont \bfseries #1}}
\newcommand{\matrixEl}[3]{{#1}_{{#2},{#3}}}
\newcommand{\matI}{\matrixSym{I}}
\newcommand{\matH}{\matrixSym{H}}
\newcommand{\matQ}{\matrixSym{Q}}
\newcommand{\matR}{\matrixSym{R}}
\newcommand{\matElR}[2]{\matrixEl{R}{#1}{#2}}
\newcommand{\vectorSym}[1]{\textbf{#1}}
\newcommand{\vectorEl}[2]{{#1}_{#2}}
\newcommand{\vectS}{\vectorSym{s}}
\newcommand{\vectElS}[1]{\vectorEl{s}{#1}}
\newcommand{\vectElSEnum}[2]{\vectorEl{s}{#1}^{(#2)}}
\newcommand{\vectElSA}[1]{\vectElS{\mathtext{A},#1}}
\newcommand{\vectElSAEnum}[2]{\vectElSA{#1}^{(#2)}}
\newcommand{\vectElSC}[1]{\vectElS{\mathtext{C},#1}}
\newcommand{\vectElSCEnum}[2]{\vectElSC{#1}^{(#2)}}
\newcommand{\Es}{E_\mathtext{s}}
\newcommand{\vectY}{\vectorSym{y}}
\newcommand{\vectElY}[1]{\vectorEl{y}{#1}}
\newcommand{\vectN}{\vectorSym{n}}
\newcommand{\real}[1]{\mathbb{R}^{#1}}
\newcommand{\complex}[1]{\mathbb{C}^{#1}}
\newcommand{\realpart}[1]{\mathrm{Re}\lbrace#1\rbrace}
\newcommand{\imagpart}[1]{\mathrm{Im}\lbrace#1\rbrace}
\newcommand{\Mt}{M_{\mathtext{T}}}
\newcommand{\Mr}{M_{\mathtext{R}}}
\newcommand{\constellation}{\mathcal{O}}
\newcommand{\constellationvector}{\mathcal{O}^{\Mt}}
\newcommand{\bps}{Q}
\newcommand{\Nzero}{N_\mathtext{0}}
\newcommand{\yQR}{\tilde{\vectY}}
\newcommand{\yQREl}[1]{\tilde{\vectElY{#1}}}
\newcommand{\nQR}{\tilde{\vectN}}
\newcommand{\syms}[1]{\chi^{(#1)}_{i,b}}
\newcommand{\mymin}[2]{\underset{#1}{\min}\left\lbrace#2\right\rbrace}
\newcommand{\myargmin}[2]{\underset{#1}{\arg \min}\left\lbrace#2\right\rbrace}
\newcommand{\bit}{\bitvar{i,b}}
\newcommand{\bitvar}[1]{x_{#1}}
\newcommand{\dibVar}[1]{d_{i,#1}}
\newcommand{\dib}{\dibVar{b}}
\newcommand{\di}{d_{i}}
\newcommand{\diEnum}[1]{d_{i}^{(#1)}}
\newcommand{\xmapVar}[1]{\bit^{\mathtext{MAP}#1}}
\newcommand{\xmapVAR}[2]{\bitvar{#2}^{\mathtext{MAP}#1}}
\newcommand{\xmap}{\xmapVar{}}
\newcommand{\xmapOld}{\xmapVar{,\mathtext{old}}}
\newcommand{\xmapCur}{\xmapVar{,\mathtext{cur}}}
\newcommand{\xmapCurVAR}[1]{\xmapVAR{,\mathtext{cur}}{#1}}
\newcommand{\smap}{\vectS^\mathtext{MAP}}
\newcommand{\xnmap}{\overline{\bit^\mathtext{MAP}}}
\newcommand{\xnmapVar}[1]{\overline{\bitvar{#1}^\mathtext{MAP}}}
\newcommand{\LE}{L^\mathtext{E}_{i,b}}
\newcommand{\LEclp}{L^\mathtext{E}_{i,b,\mathtext{clipped}}}
\newcommand{\LD}{L^\mathtext{D}_{i,b}}
\newcommand{\LA}{\LAvar{i,b}}
\newcommand{\LAvar}[1]{L^\mathtext{A}_{#1}}
\newcommand{\lmax}{L^\mathtext{E}_\mathtext{max}}
\newcommand{\metric}{\lambda}
\newcommand{\lmapvar}[1]{\metric^{\mathtext{MAP}#1}}
\newcommand{\lmap}{\lmapvar{}}
\newcommand{\lmapOld}{\lmapvar{,\mathtext{old}}}
\newcommand{\lmapCur}{\lmapvar{,\mathtext{cur}}}
\newcommand{\lnmapvar}[1]{\metric^{\overline{\mathtext{MAP}}#1}_{i,b}}
\newcommand{\lnmapVAR}[2]{\metric^{\overline{\mathtext{MAP}}#1}_{#2}}
\newcommand{\lnmap}{\lnmapvar{}}
\newcommand{\lnmapCur}{\lnmapvar{,\mathtext{cur}}}
\newcommand{\lnmapCurVAR}[1]{\lnmapVAR{,\mathtext{cur}}{#1}}
\newcommand{\metricExtr}{\Lambda}
\newcommand{\lmapExtrVar}[1]{\metricExtr^{\mathtext{MAP}#1}}
\newcommand{\lmapExtr}{\lmapExtrVar{}}
\newcommand{\lmapOldExtr}{\lmapExtrVar{,\mathtext{old}}}
\newcommand{\lnmapExtrVar}[2]{\metricExtr^{\overline{\mathtext{MAP}}#1}_{i,b#2}}
\newcommand{\lnmapExtrVAR}[2]{\metricExtr^{\overline{\mathtext{MAP}}#1}_{#2}}
\newcommand{\lnmapExtr}{\lnmapExtrVar{}{}}
\newcommand{\lnmapExtrCur}{\lnmapExtrVar{,\mathtext{cur}}{}}
\newcommand{\lnmapExtrClp}{\lnmapExtrVar{}{,\mathtext{clipped}}}
\newcommand{\numExaminedNodes}{N_{\mathtext{en}}}
\newcommand{\legacyNumVisitedNodes}{N_{\mathtext{vn}}}
\newcommand{\avgExaminedNodes}{\mathbb{E}[\numExaminedNodes]}
\newcommand{\legacyAvgNumVisitedNodes}{\mathbb{E}[\legacyNumVisitedNodes]}
\newcommand{\Mp}[1]{\mathcal{M}_{\mathtext{P}}#1}
\newcommand{\Mc}[1]{\mathcal{M}_{\mathtext{C}}#1}
\newcommand{\Ma}[1]{\mathcal{M}_{\mathtext{A}}#1}
\newcommand{\MaMin}[1]{\mathcal{M}_{\mathtext{A,min}}#1}
\newcommand{\MpruningSameLevel}[1]{\mathcal{M}_{\textrm{prn},#1}^{\textrm{sibl.}}}
\newcommand{\MpruningStepDown}[1]{\mathcal{M}_{\textrm{prn},#1}^{\textrm{down}}}
\newcommand{\mapper}{\mathcal{M}}
\newcommand{\demapper}{\mathcal{D}}
\newcommand{\quantizer}{\mathcal{Q}}
\newcommand{\rate}{r}
\newcommand{\throughput}{\Theta}
\newcommand{\clkFreq}{f_{\mathtext{clk}}}
\newcommand{\sign}[1]{\mathrm{sign}(#1)}
\newcommand{\blockNumChannelEnumeration}{\ding{172}}
\newcommand{\blockNumAprioriEnumeration}{\ding{173}}
\newcommand{\blockNumPruning}{\ding{174}}
\newcommand{\blockNumMPHistory}{\ding{175}}
\newcommand{\blockNumPreferredSiblings}{\ding{176}}
\newcommand{\blockNumColZigzag}{\ding{177}}
\newcommand{\blockNumEnumeratedNodes}{\ding{178}}
\newcommand{\blockNumAprioriMetrics}{\ding{179}}
\newcommand{\blockNumAprioriMinSearch}{\ding{180}}
\newcommand{\setA}{\mathcal{A}\left(\mathbf{x}^{(i)}\right)}
\newcommand{\setAOne}{\mathcal{A}_1\left(\mathbf{x}^{(i)}\right)}
\newcommand{\setATwo}{\mathcal{A}_2\left(\mathbf{x}^{(i)}\right)}
\newcommand{\setB}{\mathcal{B}\left(\mathbf{x}^{(i)}\right)}
\newcommand{\setBOne}{\mathcal{B}_1\left(\mathbf{x}^{(i)}\right)}
\newcommand{\setBTwo}{\mathcal{B}_2\left(\mathbf{x}^{(i)}\right)}

%
%
%
%

\newcommand{\mytitle}{A Scalable VLSI Architecture for Soft-Input Soft-Output Depth-First Sphere Decoding${}^{*}$
}

\newcommand{\mykeywords}{VLSI architecture,
                         Schnorr-Euchner (SE) enumeration,
                         iterative multiple-input multiple-output (MIMO) decoding,
                         \softinput{} \softoutput{} (SISO) sphere decoding (SD)}

\title{\mytitle}

\author{
    Ernst Martin Witte,
    Filippo Borlenghi,
    Gerd Ascheid, {\it Senior IEEE},
    Rainer Leupers,
    Heinrich Meyr, {\it Fellow IEEE}
}

\maketitle
\thispagestyle{fancy}

%
%
%
%

\begin{abstract}
Multiple-input multiple-output (MIMO) wireless transmission imposes huge
challenges on the design of efficient hardware architectures for iterative receivers. A major
challenge is  \softinput{} \softoutput{} (SISO) MIMO demapping, often approached by sphere
decoding (SD).
In this paper, we introduce the---to our best knowledge---first VLSI architecture for SISO SD applying a
single tree-search approach. Compared with a \softoutputOnly{} base architecture similar to the
one proposed by Studer et al. in IEEE J-SAC 2008, the architectural modifications for soft input
still allow a one-node-per-cycle execution. For a 4$\times$4 16-QAM system, the area increases by  \unit[57]{\%}
and the operating frequency degrades by \unit[34]{\%} only.

\end{abstract}

%
%
%
%
%
%

\begin{figure}[b]
\begin{minipage}{\columnwidth}
\setlength{\parindent}{2ex}
\footnotesize{
Manuscript received October 26, 2009; revised April 5, 2010.
This work has been supported by the UMIC (Ultra High-Speed Mobile Information and Communication)
Research Centre at the RWTH-Aachen University.

The authors are with the Institute for Integrated Signal Processing Systems, RWTH-Aachen University,
D-52056 Aachen, Germany (email: \{witte,borlenghi,ascheid,leupers,meyr\}@iss.rwth-aachen.de).
}
\end{minipage}
\end{figure}

%
%
%
%


%
%
%
%

\section{Introduction}

Multiple-input multiple-output (MIMO) wireless transmissions utilizing spatial multiplexing
achieve an increased spectral efficiency compared with single-antenna systems. This
improvement comes at the cost of an increased signal-demapping complexity, which becomes particularly
critical for iterative receivers \cite{Hochwald2003}.
Recent developments of \softinput{} \softoutput{} (SISO) MIMO-demapping algorithms reduced this
complexity significantly. Prominent demapping algorithms are  k-best and list-based approaches
\cite{Chen2007a, Li2008a}, Markov chain Monte Carlo algorithms (MCMC) \cite{Laraway2009}
 and single tree-search (STS) sphere decoders (SD)
\cite{Studer2009-arxiv}. The STS approach is often preferred since it guarantees max-log maximum a posteriori (MAP)
optimality.

Efficient VLSI implementations have been proposed for \softoutputOnly{} \STSSD{}s
\cite{Studer2008, Mennenga2009} exploiting geometric properties of QAM constellations.
These geometric relations help determining a search order, defined as
\emph{enumeration}, leading to a fast average tree-search convergence.
The SISO STS complexity has been prohibitive for VLSI
implementations so far, because geometric relations are not applicable directly.
Recent improvements of \softinput{} enumeration strategies moved \SISOSTSSD{} closer to VLSI
architectures \cite{Liao2009}.

\emph{Contributions:} In this paper, we introduce the---to our best knowledge---first VLSI
architecture for \SISOSTSSD. It is based on a \softoutputOnly{} architecture
following the one-node-per-cycle (ONPC) paradigm used by \cite{Studer2008}.
The SISO modifications are modular enough to be applied to other existing \STSSD{} architectures
and still allow ONPC execution. Compared with a \softoutputOnly{} architecture, the area increases by \unit[57]\% and the clock
frequency degrades by \unit[34]{\%} for a $4\times4$ 16-QAM system. Thus, this architecture enables STS-based iterative wireless MIMO receivers.

The paper is organized as follows: Section \ref{sec:siso-sphere-decoding}
sums up the basics of \SISOSTSSD{}, extended by the \softinput{} enumeration strategy in Section
\ref{sec:hybrid-enumeration}. Section \ref{sec:vlsi-architecture} describes important implementation
aspects of the scalable VLSI architecture. In Section~\ref{sec:results} the parameter design space of the
SISO STS architecture as well as area, timing and throughput are discussed.

%
%
%
%

\section{Single Tree-Search \SoftInput{} Sphere Decoding}
\label{sec:siso-sphere-decoding}

A spatial-multiplexing MIMO scheme with $\Mt$ transmit and $\Mr \ge \Mt$ receive antennas
is assumed~\cite{Hochwald2003}. Each transmit antenna sends one of the
$2^Q$ complex elements of the symbol set $\constellation$ defined by the modulation alphabet, which is assumed
to be the same for every antenna. Each vector $\vectS = \left[ s_1, ...,
s_{\Mt}\right]^T \in \constellationvector$ results from mapping $\Mt{}Q$ bits $\bit \in
\lbrace+1, -1\rbrace$ to an element of $\constellationvector$, with $i$ being the antenna index and
$b$ the bit index for one scalar symbol $s_i$.

The received symbol vector $\vectY \in \complex{\Mr}$ is given by $\vectY~=~\matH\vectS~+~\vectN$,
where $\matH \in \complex{\Mr \times \Mt}$ is the channel matrix and $\vectN \in \complex{\Mr}$ is a white
circular Gaussian noise vector with variance $\Nzero$ per element. For tree-search SD,
$\matH$ is typically QR-decomposed (QRD) with $\matH =
\matQ\matR$, $\matQ \in \complex{\Mr \times \Mt} $ and $\matQ^H\matQ=\matI$
and $\matR \in \complex{\Mt \times \Mt}$ being an upper triangular matrix \cite{Hochwald2003,
Studer2009-arxiv}. With $\yQR = \matQ^H\vectY$ and $\nQR = \matQ^H\vectN$, this results in
\begin{equation}
 \yQR = \matR\vectS + \nQR \label{eqn:transmission_model} \quad .
\end{equation}

According to \cite{Studer2009-arxiv}, the triangular matrix $\matR$ in equation (\ref{eqn:transmission_model}) allows to formulate the
SISO max-log MAP MIMO detection problem as STS within a $2^Q$-ary complete tree.
The tree levels correspond to the $\Mt$ antennas, each node $\vectElS{i} \in \constellation$ on tree level $i$ is
a received symbol candidate, with $s_1$ being a leaf node. An exhaustive search in such a tree leads
to a worst-case run-time complexity of $O(2^{\bps\Mt})$.
As formalized in equations (\ref{eqn:MA-definition}) to (\ref{eqn:MP-local-definition}),
metric increments $\Mc{(\vectElS{i})}$ for channel-based and $\Ma{(\vectElS{i})}$ for \aprioribased{}
information are summed up to a total increment $\Mp{(\vectElS{i})}$. $\probab{\vectElS{i}}$ is the
symbol probability computed from the \apriori{} log-likelihood ratios (LLRs) $\LA$.
\begin{eqnarray}
  \Ma{(\vectElS{i})}   &=& -\log \probab{\vectElS{i}}                                                        \label{eqn:MA-definition}        \\
  \Mc{(\vectElS{i})}   &=& \frac{1}{\Nzero} | \yQREl{i} - \sum_{j=i}^{\Mt}\matElR{i}{j} \vectElS{j} |^2    \label{eqn:MC-definition}        \\
  \Mp{(\vectElS{i})}   &=& \Mc{(\vectElS{i})} + \Ma{(\vectElS{i})}                                           \label{eqn:MP-local-definition}
\end{eqnarray}
The sum of metric increments along a path from the root to
node $\vectElS{i}$ yields the partial metric $\Mp{(\vectS^{(i)})}$ for a partial symbol vector
$\vectS^{(i)} = [s_i,...,s_{\Mt}]^T$:
\begin{eqnarray}
  \Mp{(\vectS^{(i)})}  &=& \sum_{j=i}^{\Mt} \Mp{(\vectElS{j})}                                               \label{eqn:MP-full-definition}
\end{eqnarray}
During a STS, the MAP solution $\smap$, its bits $\xmap$ and metric $\lmap\!=\!\Mp{(\smap)}$  and
extrinsic \counterHypothesis{} metrics $\lnmapExtr$ are computed by successively improving the
current metrics $\lmapCur$ and $\lnmapExtrCur$.
$\LE$ are extrinsic LLRs with
\begin{eqnarray*}
  \smap                &=& \myargmin{\vectS \in \constellationvector}{\Mp{(\vectS)}}            \\
  \lnmapExtr           &=& \mymin{\vectS \in \constellationvector \wedge \bit \ne \xmap}{\Mp{(\vectS)}} - \LA \xmap \\
  \LE                  &=& \left(\lnmapExtr - \lmap\right) \xmap \quad .
\end{eqnarray*}

These metric computations dominate the detection
complexity.
For a depth-first tree search, the pruning of \subtree{}s lying outside a hypersphere with a radius
not improving
$\lnmapCur\!=\!\lnmapExtrCur\!+\!\LA \xmap$
provides a heuristic for complexity reduction which is sensitive to
the visiting order $[\vectElSEnum{i}{1}, ..., \vectElSEnum{i}{|\constellation|}]$.
A Schnorr-Euchner (SE) order \cite{Schnorr1994} provides a very fast search
convergence by the following pruning criteria \cite{Studer2008}, typically defining the pruning
metrics $\MpruningStepDown{j}\!:=\!\MpruningSameLevel{j}\!:=\!\Mp{(\vectS^{(j)})}$:
\begin{eqnarray}
    \MpruningStepDown{j}  \ge \max{ \left \lbrace \left. \lnmapCur \right| i <   j \vee \bit \ne \xmapCur, \forall b \right\rbrace }
            \label{eqn:criterion-step-down} \\
    \MpruningSameLevel{j} \ge \max{ \left \lbrace \left. \lnmapCur \right| i \le j \vee \bit \ne \xmapCur, \forall b \right\rbrace }
            \label{eqn:criterion-step-sibling}
\end{eqnarray}
If inequality~(\ref{eqn:criterion-step-down}) holds, the current node and its \subtree{} are pruned,
otherwise a step down is performed in the tree.
If inequality~(\ref{eqn:criterion-step-sibling}) holds, the enumeration on level $j$ stops,
otherwise the sibling of the current node is enumerated.
The arguments of the $\max{}$ operators in~(\ref{eqn:criterion-step-down}) and~(\ref{eqn:criterion-step-sibling}) are
the sets $\mathcal{A}$ and $\mathcal{B}$ respectively in~\cite{Studer2008}.
We define an \emph{examined node} (as used in \cite{Studer2008} and \cite{Mennenga2009}) as a
node $\vectElS{j}$ that has been checked against at least one pruning criterion, leading to the complexity measure
\emph{number of examined nodes per detected symbol vector} $\numExaminedNodes$.

If a leaf node with $\Mp{(\vectS)} \ge \lmapCur$ is not pruned
by inequalities~(\ref{eqn:criterion-step-down}) or (\ref{eqn:criterion-step-sibling}),
the values $\lbrace\lnmapExtrCur|\bit \ne \xmapCur\rbrace$ need to be
updated by $\min{\lbrace \lnmapExtrCur, \Mp{(\vectS)} - \LA\xmapCur \rbrace}$.
Otherwise, if $\Mp{(\vectS)} < \lmapCur$, the current leaf becomes the new MAP solution and the extrinsic \counterHypothesis{} metrics $\lbrace \lnmapExtrCur |
\xmapOld \ne \xmapCur \rbrace$ are updated by $\min{\lbrace \lnmapExtrCur,
\lmapOld - \LA\xmapCur \rbrace}$.

Many methods exist to reduce $\numExaminedNodes$, like sorted QRD (SQRD)
\cite{Wubben2001} and extrinsic LLR clipping \cite{Studer2009-arxiv}. The latter one
limits the allowed range for $\LE$ to $|\LEclp| \le \lmax$, which leads to
clipped extrinsic metrics $\lnmapExtrClp$:
\begin{equation}
  \lnmapExtrClp\!=\!\max\left\lbrace \lmap\!-\!\lmax, \min\left\lbrace \lmap\!+\!\lmax, \lnmapExtr \right\rbrace \right\rbrace
  \label{eqn:llr-clipping}
\end{equation}
Please note that equation (\ref{eqn:llr-clipping}) is stricter than the $\min\{\}$ function used in \cite{Studer2009-arxiv}  where
a post-processing step is used to guarantee $|\LEclp|~\le~\lmax$ for proper channel decoding. In \cite{Studer2009-arxiv}, this saves \unit[50]{\%} of the comparisons required for clipping.
Experiments indicate that $\avgExaminedNodes$
differs only marginally between the two clipping methods.
Moreover, radius tightening further reduces $\numExaminedNodes$.
A hardware-friendly approximation of $\Ma{(\vectElS{i})}$ for statistically independent symbols, including tightening and still
guaranteeing max-log-optimal a posteriori LLRs, has been proposed in \cite{Studer2009-arxiv} (with unipolar bits $\dib\!=\!\frac{1}{2}(1-\bit\cdot\sign{\LA})$):
\begin{equation}
    \Ma{(\vectElS{i})} = -\log \probab{\vectElS{i}} \approx \sum_{b=1}^{Q} \begin{cases}
                                                                          |\LA|, & \dib = 1 \\
                                                                          0,     & \textrm{otherwise}
                                                                      \end{cases}
    \label{eqn:symbol-probability-approximation}
\end{equation}

%
%
%
%

\section{The Hybrid-Enumeration Algorithm}
\label{sec:hybrid-enumeration}

A major issue of SD algorithms is the \emph{enumeration} process, namely the
determination of the SE order $[\vectElSEnum{i}{1}, ..., \vectElSEnum{i}{|\constellation|}]$
on a level $i$ with $\vectElSEnum{i}{k}$ representing the $k^{\mathtext{th}}$ candidate for node
$\vectElS{i}$, in ascending order of $\Mp$.
A straightforward implementation by computing and fully
sorting the set $\lbrace\Mp{(\vectElSEnum{i}{k})}\rbrace$ is very expensive and inefficient.
For the \softoutputOnly{} case, the geometric properties of the QAM constellation can be exploited to avoid full sorting and
thus save most of the computations, as proposed in \cite{Studer2008, Hess2007, Mennenga2009}.
However, in iterative receivers these optimizations are not usable directly because the geometry-based order is
scrambled by the \apriori{} information.
A viable approach towards efficient \softinput{} enumeration
is given by the \emph{hybrid-enumeration} algorithm presented in \cite{Liao2009}. Its basic idea
is to split the enumeration of $\lbrace\Mp{(\vectElSEnum{i}{k})}\rbrace$ into two concurrent enumerations of $\lbrace\Mc{(\vectElSEnum{i}{k})}\rbrace$
and $\lbrace\Ma{(\vectElSEnum{i}{k})}\rbrace$.

On the one hand, the enumeration of $\lbrace\Mc{(\vectElSEnum{i}{k})}\rbrace$ is the same as in the \softoutputOnly{} case,
thus allowing to reuse any of the related aforementioned efficient methods, even in later iterations. On the other hand,
the enumeration of $\lbrace\Ma{(\vectElSEnum{i}{k})}\rbrace$ is efficient as well since the
linear sorting of the symbol set $\constellation$ needs to be performed independently only once per antenna.

According to \cite{Liao2009}, the channel- and \aprioribased{} enumerations independently select candidate
symbols $\vectElSCEnum{i}{k}$ and $\vectElSAEnum{i}{k}$ at each step $k$. The hybrid enumeration
simply selects the candidate with the lower metric $\Mp$ between these two.

\begin{figure}
    \begin{center}
        \psfrag{Ma}[Bc][Bc][1][0]{\footnotesize $\Ma$}
        \psfrag{Mc}[Br][Br][1][0]{\footnotesize $\Mc$}

        \psfrag{step}[Bc][Bc][1][0]{\scriptsize step $k$}
        \psfrag{k1}[Bc][Bc][1][0]{\scriptsize $1$}
        \psfrag{k2}[Bc][Bc][1][0]{\scriptsize $2$}
        \psfrag{k3}[Bc][Bc][1][0]{\scriptsize $3$}
        \psfrag{k4}[Bc][Bc][1][0]{\scriptsize $4$}

        \psfrag{CMa}[Bc][Bc][1][0]{\scriptsize $\Mp{(\vectElSAEnum{i}{k})}$}
        \psfrag{ma1}[Bc][Bc][1][0]{\scriptsize $\Mp{(O^{(2)})}$}
        \psfrag{ma2}[Bc][Bc][1][0]{\scriptsize $\Mp{(O^{(1)})}$}
        \psfrag{ma3}[Bc][Bc][1][0]{\scriptsize $\Mp{(O^{(4)})}$}
        \psfrag{ma4}[Bc][Bc][1][0]{\scriptsize \textcolor{gray}{$\Mp{(O^{(4)})}$}}
        \psfrag{ma5}[Bc][Bc][1][0]{\scriptsize \textcolor{gray}{$\Mp{(O^{(4)})}$}}
        \psfrag{ma6}[Bc][Bc][1][0]{\scriptsize $\Mp{(O^{(4)})}$}

        \psfrag{r1} [Bc][Bc][1][0]{\scriptsize $<$}
        \psfrag{r2} [Bc][Bc][1][0]{\scriptsize $<$}
        \psfrag{r3} [Bc][Bc][1][0]{\scriptsize $>$}
        \psfrag{r4} [Bc][Bc][1][0]{\scriptsize \textcolor{gray}{$>$}}
        \psfrag{r5} [Bc][Bc][1][0]{\scriptsize \textcolor{gray}{$>$}}
        \psfrag{r6} [Bc][Bc][1][0]{\scriptsize $=$}

        \psfrag{CMc}[Bc][Bc][1][0]{\scriptsize $\Mp{(\vectElSCEnum{i}{k})}$}
        \psfrag{mc1}[Bc][Bc][1][0]{\scriptsize $\Mp{(O^{(3)})}$}
        \psfrag{mc2}[Bc][Bc][1][0]{\scriptsize $\Mp{(O^{(3)})}$}
        \psfrag{mc3}[Bc][Bc][1][0]{\scriptsize $\Mp{(O^{(3)})}$}
        \psfrag{mc4}[Bc][Bc][1][0]{\scriptsize \textcolor{gray}{$\Mp{(O^{(1)})}$}}
        \psfrag{mc5}[Bc][Bc][1][0]{\scriptsize \textcolor{gray}{$\Mp{(O^{(2)})}$}}
        \psfrag{mc6}[Bc][Bc][1][0]{\scriptsize $\Mp{(O^{(4)})}$}

        \psfrag{E}  [Bc][Bc][1][0]{\scriptsize $\vectElSEnum{i}{k}$}
        \psfrag{n1} [Bc][Bc][1][0]{\scriptsize $O^{(2)}$}
        \psfrag{n2} [Bc][Bc][1][0]{\scriptsize $O^{(1)}$}
        \psfrag{n3} [Bc][Bc][1][0]{\scriptsize $O^{(3)}$}
        \psfrag{n4} [Bc][Bc][1][0]{\scriptsize \textcolor{gray}{ \emph{skipped}} }
        \psfrag{n5} [Bc][Bc][1][0]{\scriptsize \textcolor{gray}{ \emph{skipped}} }
        \psfrag{n6} [Bc][Bc][1][0]{\scriptsize $O^{(4)}$}

        \psfrag{Mp1}[Bc][Bc][1][0]{\scriptsize $\Mp{(O^{(1)})}$}
        \psfrag{Mp2}[Bc][Bc][1][0]{\scriptsize $\Mp{(O^{(2)})}$}
        \psfrag{Mp3}[Bc][Bc][1][0]{\scriptsize $\Mp{(O^{(3)})}$}

        \psfrag{o1} [cc][cc][1][0]{\tiny       $O^{(1)}$}
        \psfrag{o2} [cc][cc][1][0]{\tiny       $O^{(2)}$}
        \psfrag{o3} [cc][cc][1][0]{\tiny       $O^{(3)}$}
        \psfrag{o4} [cc][cc][1][0]{\tiny       $O^{(4)}$}

        \psfrag{1}  [Bc][Bc][1][0]{\scriptsize $k\!=\!1$}
        \psfrag{2}  [Bc][Bc][1][0]{\scriptsize $k\!=\!2$}
        \psfrag{34} [Bc][Bc][1][0]{\scriptsize $k\!\in\!\lbrace3,\!4\rbrace$}

        \psfrag{4}  [Bc][Bc][1][0]{\scriptsize $k\!=\!4$}
        \psfrag{123}[Br][Br][1][0]{\scriptsize $k\!\in\!\lbrace1,\!2,\!3\rbrace$}

        \includegraphics[width=1\columnwidth]{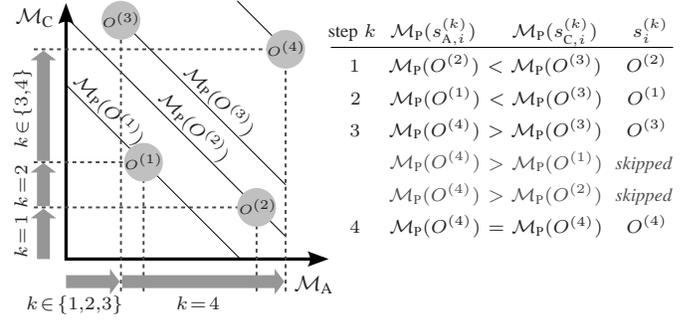}
    \end{center}
    \caption{Hybrid-enumeration example, $k^{th}$ symbol in SE order: $O^{(k)} \in \mathcal{O}$.}
    \label{fig:hybrid_enumeration}
\end{figure}

As visualized in Figure~\ref{fig:hybrid_enumeration}, the strict SE order is not preserved, hence the inequality
$\Mp{(\vectElSEnum{i}{k})} \le \Mp{(\vectElSEnum{i}{l})}, \forall l > k $ does not hold any more.
Thus, a modification of the pruning criteria is needed to avoid the erroneous
exclusion of the MAP or counter-hypothesis solutions.
For $l > k$, the inequalities
$\Mc{(\vectElSCEnum{i}{k})} \le \Mc{(\vectElSCEnum{i}{l})}$ and $\Ma{(\vectElSAEnum{i}{k})} \le \Ma{(\vectElSAEnum{i}{l})}$
lead to
$\Mc{(\vectElSCEnum{i}{k})} + \Ma{(\vectElSAEnum{i}{k})} \le \Mp{(\vectElSEnum{i}{l})}$, providing an
alternative lower bound for tree pruning.
Thus, in \cite{Liao2009} the pruning metric of inequality~(\ref{eqn:criterion-step-sibling}) on the current tree level $i$ is re-defined as
\begin{equation}
    \MpruningSameLevel{i} := \Mc{(\vectElSCEnum{i}{k})} + \Ma{(\vectElSAEnum{i}{k})} + \Mp{(\vectS^{(i+1)})} \quad.
    \label{eqn:re-defined-criterion-step-sibling}
\end{equation}
Compared with the SE order,
pruning metric~(\ref{eqn:re-defined-criterion-step-sibling}) preserves the error-rate performance
at the price of a slight increase in $\numExaminedNodes$.
For a more detailed description and analysis of the hybrid-enumeration
algorithm, the reader is referred to \cite{Liao2009}.

%
%
%
%

\section{A VLSI Architecture for STS \SoftInput{} Sphere Decoding}
\label{sec:vlsi-architecture}
In this section, a VLSI architecture for \SISOSTSSD{} is introduced.
It is derived from a \softoutputOnly{} depth-first STS base architecture
extended by \softinput{} processing.
The main challenges are discussed that arise from the implementation of efficient \softinput{} extensions
according to the hybrid-enumeration scheme.
Further algorithmic optimizations such as LLR correction proposed in \cite{Studer2009-arxiv} are orthogonal
to the base architecture and can be implemented on top of it.

\subsection{\SoftOutputOnly{} Base Architecture}
\label{sec:base-architecture}
\begin{figure}
    \begin{center}
        \psfrag{control}  [cc][cc][1][0]{STS Control FSM}
        \psfrag{InputRegs}[bl][bl][1][0]{$\matR, \yQR, \lbrace\LA\rbrace$}
        \psfrag{MatVect}  [cc][cc][1][0]{\small$\yQREl{i}\!-\!\sum\limits_{j=i+1}^{\Mt}\!\matElR{i}{j}\vectElS{j}$}
        \psfrag{Cache}    [bc][bc][1][0]{\small{$\Mt2^Q$ Enumerated Nodes Flags}}
        \psfrag{colzz}    [bc][bc][1][0]{col. \zigzag{}}
        \psfrag{MC}       [bl][bl][1][0]{$\Mc$}
        \psfrag{MA}       [bl][bl][1][0]{$\Ma$}
        \psfrag{MA0}      [bl][bl][1][0]{$\Ma = 0$}
        \psfrag{MP}       [bl][bl][1][0]{$\Mp$}
        \psfrag{M}        [bc][bc][1][0]{$\mapper$}
        \psfrag{D}        [bc][bc][1][0]{$\demapper$}
        \psfrag{Q}        [cc][cc][1][0]{$\quantizer$}
        \psfrag{cc}       [cc][cc][1][0]{Pruning Criteria, $\lbrace\lnmapExtr\rbrace, \lmap, \lbrace\xmap\rbrace$}
        \psfrag{LE}       [cc][cc][1][0]{$\lbrace\LE\rbrace$}
        \psfrag{CSd}      [cc][cc][1][0]{min}
        \psfrag{CSh}      [cc][cc][1][0]{min}
        \psfrag{min}      [cc][cc][1][0]{\small $\MaMin$}
        \psfrag{min2}     [cc][cc][1][0]{\footnotesize $2^Q : 1$}
        \psfrag{cmin}     [cc][cc][1][0]{\scriptsize{min}}
        \psfrag{prefch}   [cc][cc][1][0]{\scriptsize{$\Mt-1$ Pref. Siblings}}
        \psfrag{hist}     [cc][cc][1][0]{$\Mp$ History}

        \psfrag{MAs}      [cc][cc][1][0]{\small $\lbrace\Ma\rbrace_{i}$}

        \psfrag{c1}       [cc][cc][1][0]{\blockNumChannelEnumeration}
        \psfrag{c2}       [cc][cc][1][0]{\blockNumAprioriEnumeration}
        \psfrag{c3}       [cc][cc][1][0]{\blockNumPruning}
        \psfrag{c4}       [cc][cc][1][0]{\blockNumPreferredSiblings}
        \psfrag{c5}       [cc][cc][1][0]{\blockNumEnumeratedNodes}
        \psfrag{c6}       [cc][cc][1][0]{\blockNumColZigzag}
        \psfrag{c7}       [cc][cc][1][0]{\blockNumAprioriMetrics}
        \psfrag{c8}       [cc][cc][1][0]{\blockNumAprioriMinSearch}
        \psfrag{c9}       [cc][cc][1][0]{\blockNumMPHistory}
        \includegraphics[width=\columnwidth]{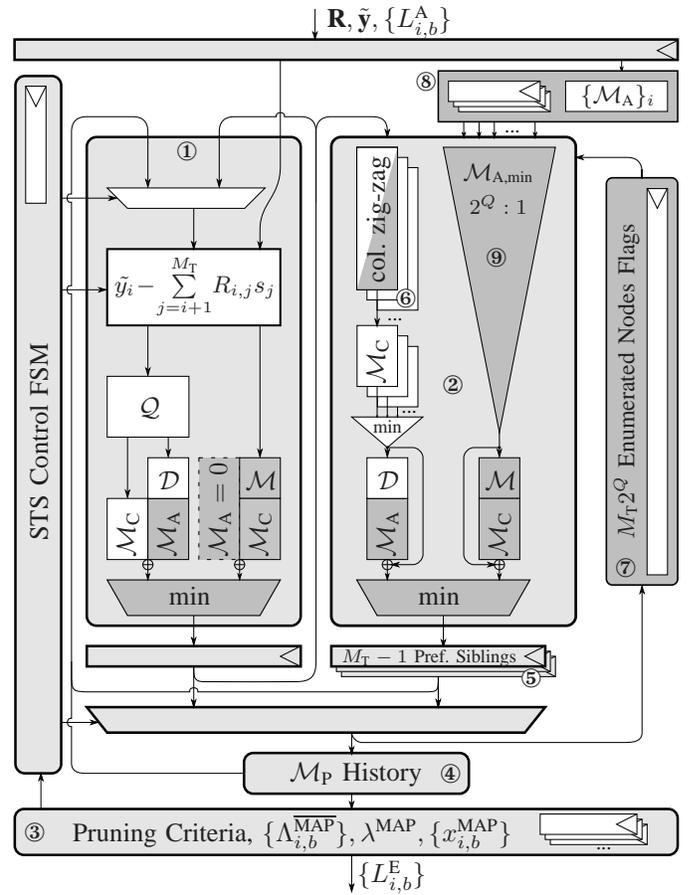}
    \end{center}
    \caption{Block diagram of the proposed \softinput{} \STSSD{} VLSI architecture.
    Units added/modified for \softinput{} are emphasized by dark gray background.
    Legend: Mapper $\mapper$, Demapper $\demapper$, Quantizer $\quantizer$.}
    \label{fig:vlsi-architecture}
\end{figure}

The \softoutputOnly{} base STS architecture, composed of the light gray blocks in
Figure~\ref{fig:vlsi-architecture},
follows the ONPC execution principle used by Studer et al{.} in \cite{Studer2008}.
Its architectural structure is derived from the observation that the
tree search is composed of three basic control-flow steps:

\emph{i) Vertical steps} (\blockNumChannelEnumeration) down from tree level $i$ to $i-1$ enumerate the first child node
$\vectElS{i-1}^{(1)}$ of    a parent node    $\vectElS{i}^{(k)}$. This requires
    a quantization step $\mathcal{Q}$ to find the QAM symbol next to $\yQREl{i}$,
        followed by the computation of $\Mp{(\vectElS{i-1}^{(1)})}$.
    The result of $\mathcal{Q}$ is used to initialize the enumeration on the tree
    level $i-1$ and by the pruning-criteria check for $\vectElS{i-1}^{(1)}$.

\emph{ii) Horizontal steps} (\blockNumAprioriEnumeration) on a tree level $i$ enumerate the node $\vectElS{i}^{(k+1)}$ after
    enumerating the node  $\vectElS{i}^{(k)}$ and its \subtree{}. This category also includes steps back from a child
    node    $\vectElS{i-1}$ to the next sibling $\vectElS{i}^{(k+1)}$ of its parent
    node $\vectElS{i}^{(k)}$.

\emph{iii) Pruning-criteria checks} (\blockNumPruning) for a node $\vectElS{i}^{(k)}$ determine if either a
    \emph{vertical step} to the child $\vectElS{i-1}^{(1)}$, a \emph{horizontal step} to the sibling
    $\vectElS{i}^{(k+1)}$ or a \emph{horizontal step} to its parent's sibling
    $\vectElS{i+1}^{(l+1)}$ has to be performed next.
    The \emph{$\Mp$ history} (\blockNumMPHistory) unit stores the partial metrics
    $\Mp{(\vectS^{(i)})}$, recursively implements equation (\ref{eqn:MP-full-definition})
    and provides its result to unit \blockNumPruning{} for pruning and LLR clipping by equation (\ref{eqn:llr-clipping}).

In a depth-first SD, the tree-traversal control flow exhibits severe data and control
dependencies. In order to achieve a throughput of one examined node per cycle, the base architecture
executes the pruning check for node $\vectElS{i}^{(k)}$ concurrently with
the steps towards $\vectElSEnum{i-1}{1}$ and $\vectElSEnum{i}{k+1}$ in cycle $n$. If the
pruning check selects $\vectElSEnum{i-1}{1}$, $\vectElSEnum{i}{k+1}$ is saved in a
\emph{preferred-siblings cache} (\blockNumPreferredSiblings) for later use during a step up in the tree.
Thus, in cycle $n+1$ the availability of a valid node for the next pruning check is guaranteed.

The enumeration unit of the base architecture employs the \colwise{} \zigzag{} enumeration strategy (\blockNumColZigzag)
presented in \cite{Hess2007}. Compared with circular PSK-like enumeration \cite{Studer2008},
the \colwise{} enumeration allows a much more regular hardware implementation. Furthermore, for 64 QAM and
higher modulation orders it requires less comparisons.

Since there is no assumption on the mapping between QAM symbols and bits, two run-time-programmable lookup tables,
named mapper $\mapper$ and demapper $\demapper$ respectively, are used for the conversion between the symbol and
the bit representations.

\subsection{\SoftInput{} Extensions}
\label{sec:soft-input-architecture-extensions}
In order to extend the base architecture presented in Section~\ref{sec:base-architecture}, mainly extra units for the
\aprioribased{} enumeration have to be added, along with slight changes in the \colwise{} \zigzag{} implementation. These extensions
correspond to the dark gray units in Figure~\ref{fig:vlsi-architecture}.

\subsubsection{Enumerated-nodes flags}
Both channel- and \aprioribased{} enumeration units have to skip nodes
that have already been enumerated, because the local enumeration orders for $\Mc$ and $\Ma$
differ from the global enumeration order.
Therefore, both units need the list of enumerated nodes to guarantee
that each node is enumerated only once. This flag vector of $2^Q$ bits per antenna is maintained in unit \blockNumEnumeratedNodes.

\subsubsection{Modified \colwise{} zig-zag enumeration}
Skipping an arbitrary number of nodes implies modifications to the \colwise{} \zigzag{} implementation (\blockNumColZigzag).
Compared with the base architecture, the new column-enumeration unit
does not keep internal \zigzag{} states any more. Instead, each column enumeration performs a minimum search over the linear distances
between the quantized imaginary part $\mathcal{Q}(\imagpart{\yQREl{i}-\sum_{j=i+1}^{\Mt}\matElR{i}{j}\vectElS{j}})$ and
all rows $\lbrace\imagpart{\vectElS{i}|\vectElS{i}\!\in\!\constellation}\rbrace$ masked by
the enumerated-nodes flags.
The hardware complexity increases only moderately,
because distance computations are the same for all columns and operate on words of only $Q/2 + 1$ bits.

\subsubsection{\Aprioribased{} enumeration}
\label{sec:apriori-based-enumeration-hw}

With $\di$ being the decimal representation of the bit vector $[\dibVar{\bps},...,\dibVar{1}]$,
a mapping of $\di$ to the corresponding symbol $\vectElS{i}(\di)$,
$\Ma{(\di)} = \Ma{(\vectElS{i}(\di))}$ and an order defined by $\vectElS{i}(\diEnum{k}) = \vectElSAEnum{i}{k}$,
one problem of enumerating $\lbrace\Ma\rbrace_i = \lbrace\Ma{(\di)}|0\le\di<2^{\bps}\rbrace$
is the lack of relations among a priori LLRs. Thus, the only known
solution is the full computation and sorting of $\lbrace\Ma\rbrace_i$.

First, the computation of $\lbrace\Ma\rbrace_i$ (\blockNumAprioriMetrics) requires $2^{\bps}~-~\bps~-~1$ additions per antenna and received vector. Due to
the ONPC principle and the structure of (\ref{eqn:symbol-probability-approximation}),
the number of hardware adders can be reduced by resource sharing.
The first enumeration step always
results in $\diEnum{1}=0$ and $\Ma{(\diEnum{1})}=0$, thus the subset $\lbrace\Ma\rbrace_{i,\mathtext{L}} = \lbrace\Ma{(\di)}|1\le\di\le2^{\bps-1}\rbrace$
can be computed concurrently.
In the second
step, $\Ma{(\diEnum{2})}=\min_{\forall b}{|\LA|}$ can be enumerated since $\Ma{(\diEnum{2})} \in \lbrace\Ma\rbrace_{i,\mathtext{L}}$, while the
subset $\lbrace\Ma\rbrace_{i,\mathtext{H}} = \lbrace\Ma{(\di)}|2^{\bps-1} < \di < 2^{\bps}\rbrace$ can be computed.
This approach only requires $2^{\bps-1}-1$ adders independently from $\Mt$, yielding adder savings of \unit[36]{\%} for 16 QAM and \unit[45]{\%} for 64 QAM.
Furthermore, for an ONPC architecture, no latency is added since the subsets $\lbrace\Ma\rbrace_{i,\mathtext{L}}$ and $\lbrace\Ma\rbrace_{i,\mathtext{H}}$
can be computed during the enumeration of $\vectElSAEnum{i}{1}$ and $\vectElSAEnum{i}{2}$.
Further resource sharing would result in limited gains while significantly increasing irregularity.

The second issue is sorting $\lbrace\Ma\rbrace_i$.
Since latency is typically a serious issue for run-time constrained depth-first
SD, an approach has been chosen that does not add latency for the sorting of $\lbrace\Ma\rbrace_i$.
The ONPC principle allows a minimum search (\blockNumAprioriMinSearch) for
$\MaMin$ over the set $\lbrace\Ma\rbrace_i$ for the enumeration of the current antenna~$i$,
masked by the enumerated-nodes flags.
The resulting binary tree of compare-select (CS) units would dominate the critical path already for 16 QAM.

However, the properties of equation~(\ref{eqn:symbol-probability-approximation}) can be exploited
to remove almost all comparators and CS dependencies for the first three CS levels.
The principle can be explained easily by considering the removal of
the first level: for pairs of $\lbrace\Ma{(\vectElS{i}^{(k)})}, \Ma{(\vectElS{i}^{(l)})}\rbrace$ with only
one bit $\lbrace{}b | \bitvar{i,b}^{(k)}~\ne~\bitvar{i,b}^{(l)}\rbrace$ the larger metric $\Ma{(\vectElS{i}^{(\lbrace{}k,l\rbrace)})}$ is the
one with $\bitvar{i,b}^{(\lbrace{}k,l\rbrace)}~\ne~\sign{\LA}$. This kind of decision does not need any metric comparison but can be determined by
single-bit comparisons of sign bits and enumerated-nodes flags. Selecting the minimum of 4-tuples (first two CS tree levels) differing in only two bits
$\lbrace{}b_{\lbrace{}m,n\rbrace}| \bitvar{i,b_{\lbrace{}m,n\rbrace}}^{(k)}~\ne~\bitvar{i,b_{\lbrace{}m,n\rbrace}}^{(l)}\rbrace$
requires an additional comparison $|\LAvar{i,b_m}| \gtrless |\LAvar{i,b_n}|$. However, this extra comparison is
the same for all 4-tuple \subtree{}s and does not depend on intermediate results generated in the CS tree.
Therefore, the critical path is significantly reduced.
The extension to 8-tuples (first three CS tree levels) has a total of only six parallel comparators. Thus, only
one CS unit and two 8:1 multiplexers are required for 16 QAM and only seven CS units and eight 8:1 multiplexers for 64 QAM.
Compared with a full CS tree, the comparator savings are \unit[53]{\%} in total and \unit[50]{\%} in the critical path for 16 QAM and
\unit[79]{\%} in total and \unit[33]{\%} in the critical path for 64 QAM.
Extensions to higher orders than 8-tuples are possible but would result in an exponential complexity increase.

\subsubsection{Pruning-criteria checks}
\label{sec:architecture-pruning-citerion}

In \cite{Studer2008}, the checks of the pruning criteria of equations~(\ref{eqn:criterion-step-down})
and (\ref{eqn:criterion-step-sibling}) have been simplified
to a single pruning-criterion check of equation~(\ref{eqn:criterion-step-sibling}) in order to
reduce hardware complexity, at the cost of a slight increase of $\numExaminedNodes$.
For the SISO STS SD architecture proposed in this paper, the implementation of two different pruning criteria in unit \blockNumPruning{}
is mandatory to prevent a further significant increase of $\numExaminedNodes$.
In order to avoid extra delays on the critical path,
the pruning-criteria checks are not implemented as
maximum searches but as pairs of $\Mt2^Q$ fully parallel comparators $\MpruningStepDown{j}  > \lnmap$ and
$\MpruningSameLevel{j}  > \lnmap$, followed by simple bit-masking and combining.

%
%
%
%

\section{ASIC Synthesis Results}
\label{sec:results}

The architecture presented in the previous section has been implemented in VHDL including parameters for
word lengths, $\Mt$, QAM order and a switch to enable/disable \softinput{} support. A representative set of
parameter combinations has been instantiated by layout-aware gate-level synthesis\footnote{
    UMC \unit[90]{nm} standard-performance CMOS library, typical case, Synopsys Design Compiler 2009.06-sp1 in topographical mode.
}.

\begin{figure}
        \psfrag{throughput}                   [ct][cc][1][0]{\footnotesize Throughput $\throughput$ [$10^6$ information bits/sec]}
        \psfrag{SNR}                          [ct][cc][1][0]{\footnotesize Minimum $\textrm{SNR} = \Mt{}\Es/\Nzero$ [dB] for \unit[1]{\%} FER }
        \psfrag{visitednodes}                 [cB][cc][1][0]{\footnotesize Cumulative $\avgExaminedNodes$ }
        \psfrag{matlabxxxxxxxxxxxxxxxxxxI1}   [lc][lc][1][0]{\tiny $\avgExaminedNodes$, SE order (Matlab), 1 it.}
        \psfrag{matlabxxxxxxxxxxxxxxxxxxI2}   [lc][lc][1][0]{\tiny $\avgExaminedNodes$, SE order (Matlab), 2 it.}
        \psfrag{matlabxxxxxxxxxxxxxxxxxxI4}   [lc][lc][1][0]{\tiny $\avgExaminedNodes$, SE order (Matlab), 4 it.}
        \psfrag{ASICxxxxxxxxxxxxxxxxxI1}      [lc][lc][1][0]{\tiny $\avgExaminedNodes$, SISO ASIC, 1 it.}
        \psfrag{ASICxxxxxxxxxxxxxxxxxI2}      [lc][lc][1][0]{\tiny $\avgExaminedNodes$, SISO ASIC, 2 it.}
        \psfrag{ASICxxxxxxxxxxxxxxxxxI4}      [lc][lc][1][0]{\tiny $\avgExaminedNodes$, SISO ASIC, 4 it.}
        \psfrag{throughputxxxxxxxxxxxxx}      [lc][lc][1][0]{\tiny Least-Effort $\throughput$, SISO ASIC}
        \psfrag{0i8}                          [cc][cc][1][0]{\tiny 0.8}
        \psfrag{0i4}                          [cc][cc][1][0]{\tiny 0.4}
        \psfrag{0i2}                          [cc][cc][1][0]{\tiny 0.2}
        \psfrag{0i1}                          [cc][cc][1][0]{\tiny 0.1}
        \psfrag{0i05}                         [cc][cc][1][0]{\tiny 0.05}
        \psfrag{0i8 }                         [cc][cc][1][0]{\tiny 0.8}
        \psfrag{0i4 }                         [cc][cc][1][0]{\tiny 0.4}
        \psfrag{0i2 }                         [cc][cc][1][0]{\tiny 0.2}
        \psfrag{0i1 }                         [cc][cc][1][0]{\tiny 0.1}
        \psfrag{0i05 }                        [cc][cc][1][0]{\tiny 0.05}

        \psfrag{sw1}                          [cc][cc][1][0]{\footnotesize \ding{172}}
        \psfrag{sw2}                          [cc][cc][1][0]{\footnotesize \ding{173}}
        \psfrag{sw3}                          [cc][cc][1][0]{\footnotesize \ding{174}}
        \psfrag{itx1}                         [cc][cc][1][0]{\tiny 1 it.}
        \psfrag{itx2}                         [cc][cc][1][0]{\tiny 2 it.}
        \psfrag{itx3}                         [cc][cc][1][0]{\tiny 3 it.}
        \psfrag{itx4}                         [cc][cc][1][0]{\tiny 4 it.}
        \psfrag{lax}                          [lB][lB][1][0]{\tiny $\leftarrow$}
        \psfrag{rax}                          [rB][rB][1][0]{\tiny $\rightarrow$}

        \centering{\includegraphics[width=\columnwidth]{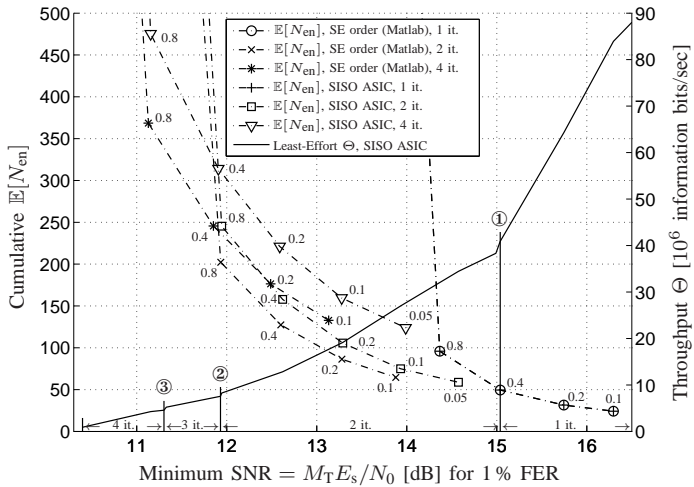}}
        \caption{Cumulative $\avgExaminedNodes$ and iterative least-effort throughput $\throughput$ over minimum SNR for \unit[1]{\%} FER for the $4\times4$ 16-QAM architecture.
                 Numbers annotated to cumulative $\avgExaminedNodes$ curves are normalized clipping levels $\Nzero\lmax$. As in \cite{Studer2009-arxiv},
                 one iteration is defined as one use of the SISO MIMO demapper and the SISO channel decoder ($1^{\mathtext{st}}$ iteration corresponds to soft-output-only SD).
        }
        \label{fig:examined_nodes}
\end{figure}

Since both the \softoutputOnly{} base architecture and the SISO architecture follow the ONPC principle,
their throughput $\throughput$ can be determined by
\begin{equation}
 \throughput = \frac{\rate \bps \Mt}{\avgExaminedNodes} \clkFreq \quad [bit/s] \label{eqn:decoding_throughput}
\end{equation}
with $\rate$ being the code rate and $\avgExaminedNodes$ being the average $\numExaminedNodes$. The curves for the iterative $\throughput$ and the cumulative $\avgExaminedNodes$ for a $4\times4$ 16-QAM MIMO
system\footnote{
    \label{footnote:system-setup}
    Throughout this paper we use a system with an i{.}i{.}d{.} Rayleigh fading
    channel, perfect channel knowledge and SQRD \cite{Wubben2001}. The BICM
    transmission is set up with a convolutional channel code (rate $1/2$, generator
    polynomials [$133_o, 171_o$], constraint length 7) decoded by a max-log BCJR channel decoder with perfect termination knowledge and an
    S-random interleaver corresponding to 512 information bits. The SNR is defined as $\textrm{SNR} = \Mt{}\Es/\Nzero$, with
    $\Es = \mathbb{E}[|s|^2], s \in \constellation$. $\probab{\vectElS{i}}$ is approximated by
    equation~(\ref{eqn:symbol-probability-approximation}).
    The VLSI architecture internally operates on normalized metrics $\mathcal{M}_\textrm{norm.}\!=\!\Nzero\mathcal{M}$ to avoid division by $\Nzero$,
    normalized clipping levels are given by $\Nzero\lmax$.
}
achieving a frame error rate (FER) of \unit[1]{\%} are given in Figure~\ref{fig:examined_nodes},
including as a reference the cumulative $\avgExaminedNodes$ obtained by
SE ordering and floating-point operations. In the $4^\textrm{th}$ iteration the hybrid-enumeration algorithm introduces
an overhead of less than \unit[28]{\%}
in terms of $\avgExaminedNodes$. 
The least-effort throughput in Figure~\ref{fig:examined_nodes} is derived from equation~(\ref{eqn:decoding_throughput})
by selecting the minimum cumulative $\avgExaminedNodes$ among all iterations for a specific SNR.
The intersections of the cumulative $\avgExaminedNodes$ curves determine the SNR points for changing the number of iterations.
In Figure~\ref{fig:examined_nodes} the switching points are marked
by \ding{172} (1 $\rightleftarrows$ 2 iterations), by \ding{173} (2 $\rightleftarrows$ 3 iterations) and by \ding{174} (3 $\rightleftarrows$ 4 iterations).

Area and delay of this architecture are quite sensitive to the fixed-point word lengths.
Therefore, the word lengths
have been carefully selected to make the FER-performance loss negligible with respect to floating-point operation\footnote{
    \label{footnote:word-lengths}
    Word lengths $[\mathtext{integer.fractional}]$ for $4\times4$ 16~QAM: $\yQREl{i}[6.7]$, $\matElR{i}{j} [4.7]$, $\LA [9.5]$, $\LE [9.5]$, $\mathcal{M}_{\lbrace\mathtext{C,A,P}\rbrace} [9.6]$.
    A QAM-order increase of factor 4 requires one more integer bit for $\yQREl{i}$ per real/imaginary part and two more integer bits
    for $\mathcal{M}_{\lbrace\mathtext{C,A,P}\rbrace}$, $\LA$ and $\LE$.
    Doubling $\Mt$ requires one more integer bit for $\mathcal{M}_{\lbrace\mathtext{C,A,P}\rbrace}$, $\LA$ and $\LE$.
}.

\begin{figure}
        \centering{\includegraphics[width=\columnwidth]{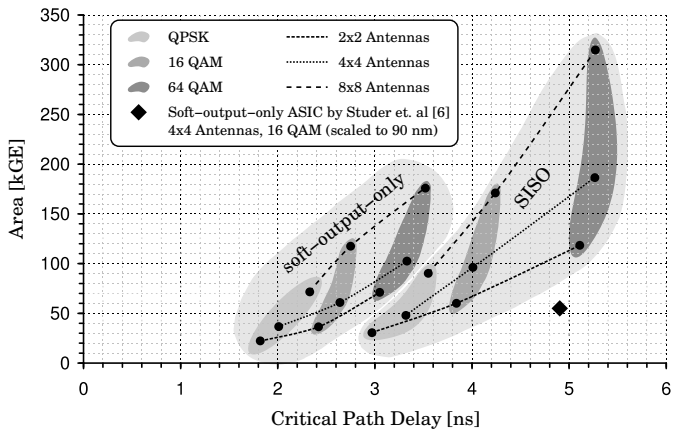}}
        \caption{Parametrization design space of the proposed STS SD architecture. Area is measured in gate equivalents (GEs). One GE corresponds
                 to the area of a two-input drive-one NAND gate.}
        \label{fig:at_chart}
\end{figure}

Figure~\ref{fig:at_chart} shows the synthesis results for representative parameter sets.
The results for the \softoutputOnly{} case
are comparable to the implementation published in~\cite{Studer2008}.
Since the two base architectures are similar, they are close in terms of area.
The timing differs, mainly for two reasons.
First, Figure~\ref{fig:at_chart} shows pre-layout synthesis results for a \unit[90]{nm} technology
whereas those in \cite{Studer2008} are post-layout results for a \unit[250]{nm} technology scaled to \unit[90]{nm} by $f_{90} \approx \frac{250}{90}f_{250}$.
Second, the architectures differ in their pipeline and enumeration schemes.

By enabling \softinput{} processing for the $4\times4$ 16-QAM reference, the area increases by \unit[57]{\%} from  \unit[61]{kGates} to \unit[96]{kGates}, while the
clock frequency degrades by \unit[34]{\%} from \unit[379]{MHz} to \unit[250]{MHz}.
We can conclude that the additional cost for \softinput{} is
affordable at the prospect of working at lower SNR regimes with iterative systems.

The proposed architecture scales almost linearly with $\Mt$ in terms of area. The critical path
degrades only by less than \unit[10]{\%} when doubling $\Mt$.
When increasing the QAM order by a factor of 4 in the \softinput{} case,
the area is less than doubled while the frequency degrades by less than
\unit[20-25]{\%}, despite the enumeration being significantly affected.

%
%
%
%

\section{Conclusion}
\label{sec:conclusion}

To our best knowledge, we introduced the first SISO STS SD architecture,
enabling iterative STS SD-based receivers. The parametrized architecture offers
very good scalability over $\Mt$ and the QAM order.
The approximate hybrid-enumeration method enables the implementation of iterative STS-based MIMO receivers,
although high data-rate communication systems may require multiple parallel SD instances to meet the throughput constraints.
We believe that the algorithms and hardware-design principles presented in this paper are suitable for
most kinds of SD architectures. Our future development will focus on
further enhancements of the architecture, based for instance on the ideas proposed in \cite{Studer2008}.

%
%
%
%

\section{Acknowledgement}

The authors would like to thank Chun-Hao Liao, I-Wei Lai, Martin
Senst, David Kammler, Andreas Minwegen, Uwe Deidersen, Konstantinos Nikitopoulos, Dan Zhang, Jeronimo Castrillon, Torsten Kempf, all
reviewers and the editor for
their valuable feedback and support.

\bibliographystyle{IEEEtran}
\bibliography{IEEEabrv,IEEEtranBST,all}
\end{document}